\documentclass[12pt,preprint]{aastex}

\newcommand{\rrab}{RR{\it ab} }
\newcommand{\rrc}{RR{\it c} }
\newcommand{\kms}{${\rm km~s}^{-1}$}

\slugcomment{Accepted for publication in the AJ}

\shorttitle{Velocity Substructures toward Virgo}
\shortauthors{Vivas et al.}

\begin{document}

\title{Spectroscopy of Bright QUEST RR Lyrae Stars: Velocity Substructures toward Virgo}

\author{A. Katherina Vivas\altaffilmark{1}, Yara L. Jaff\'e\altaffilmark{1,2},
Robert Zinn\altaffilmark{3}, Rebeccah Winnick\altaffilmark{3},
Sonia Duffau\altaffilmark{4}, Cecilia Mateu\altaffilmark{1}}

\altaffiltext{1}{Centro de Investigaciones de Astronom{\'\i}a (CIDA),
Apartado Postal 264, M\'erida 5101-A, Venezuela}
\altaffiltext{2}{School of Physics \& Astronomy, University of Nottingham, University Park, Nottingham NG7 2RD, UK}
\altaffiltext{3}{Department of Astronomy, Yale University, P.O. Box 208101,
New Haven, CT 06520-8101}
\altaffiltext{4}{Departamento de Astronom{\'\i}a, Universidad de Chile, Casilla 36-D, Santiago, Chile}

\begin{abstract}

Using a sample of 43 bright ($V<16.1$, distance $<13$ kpc) RR Lyrae
stars (RRLS) from the QUEST survey with spectroscopic radial
velocities and metallicities, we find that several separate halo
substructures contribute to the Virgo overdensity (VOD).  While there
is little evidence for halo substructure in the spatial distribution
of these stars, their distribution in radial velocity reveals two
moving groups.  These results are reinforced when the sample is
combined with a sample of blue horizontal branch stars that were
identified in the SDSS, and the combined sample provides evidence for
one additional moving group.  These groups correspond to peaks in the
radial velocity distribution of a sample of F type main-sequence stars
that was recently observed in the same directon by SEGUE, although in
one case the RRLS and F star groups may not lie at the same
distance. One of the new substructures has a very narrow range in
metallicity, which is more consistent with it being the debris from a
destroyed globular cluster than from a dwarf galaxy.  A small
concentration of stars have radial velocities that are similar to the Virgo
Stellar Stream (VSS) that was identified previously in a fainter
sample of RRLS.  Our results suggest that this feature extends to
distances as short as $\sim 12$ kpc from its previous detection at $\sim
19$ kpc.  None of the new groups and only one star in the sample have
velocities that are consistent with membership in the leading tidal
stream from the Sagittarius Dwarf Spheroidal Galaxy, which some
authors have suggested is the origin of the VOD.

\end{abstract}


\section{INTRODUCTION \label{sec-intro}}

Over the past several decades, a large number of investigations of the
stars and the globular clusters in the Galactic halo have concluded
that the halo formed chaotically and over a long period of time
through the accretion and disruption of dwarf galaxies (e.g., Searle
\& Zinn 1978, Norris 1986; Carney et al. 1990; Majewski 1992; Mackey
\& Gilmore 2004; see also Freeman \& Bland-Hawthorn 2002 for a
review).
\nocite{sz78,norris86,carney90,majewski92,mackey04,freeman02} There is
evidence, however, that the inner and outer parts of the halo have
sufficiently different properties that at least two mechanisms are
required to explain the halo
\citep[e.g.][]{hartwick87,zinn93,majewski93,chiba00,kinman07,miceli07,lee07,carollo07}.
Accretion appears to have dominated the formation of the outer halo,
but the flattening and prograde rotation of the inner halo and the old
ages of the globular clusters in the inner halo suggest that its
formation involved the dissipative merging of gas clouds at an early
epoch \citep[see][and references therein]{carollo07}.  While this dual
halo picture is supported by substantial evidence, it continues to be
important to see what fraction of the halo can be explained by, for
example, an accretion-only scenario.

The most direct evidence for accretion comes from the detection of
halo substructures consisting of co-moving streams of stars.  The best
documented of these are the streams originating in the Sagittarius
(Sgr) dwarf spheroidal (dSph) galaxy, which cross much of the sky
\citep[][among
others]{yanny00,ivezic00,ibata01a,vivas01,vivas05,newberg02,majewski03}.
Several other substructures have been recently detected in the halo of
the Milky Way \citep[][among
others]{newberg02,ibata03,majewski03,yanny03,rocha04,vivas06,belokurov06}
and in the halo of M31
\citep{ibata01b,ferguson02,morrison03,zucker04,kalirai06,fardal07}.
These streams provide strong evidence in favor of the hierarchical
model of the formation of galaxies, in which large galaxies such as
the Milky Way formed through multiple mergers with smaller building
blocks.  Simulations of the hierarchical picture predict that the
Milky Way has experienced a large number of minor mergers \citep[see
for example][]{bullock05} and show that the ones that occurred during
the past few Gyrs. may be recognizable today as stellar streams. While
these simulations appear to be consistent with the current
observations of the outer halo \citep{bell07}, the search for more
streams and better documentation of the known ones remains important.

We report here an investigation of the halo substructure in the
direction of Virgo.  The first sign of an overdensity of halo stars in
this direction came from the QUEST RR Lyrae star (RRLS) survey
\citep{vivas01}, which revealed a small overdensity near 13 hr right
ascension ($\alpha$), which was the west boundary of the survey as it
existed then.  As the survey was extended, it became clear that this
feature is considerably larger and is centered near $\alpha \sim 12.4$
hr \citep{vivas02,vivas03}. In the jargon of the QUEST survey it
became known as the {\it "12.4 hr clump"} (declination ($\delta$) was
not specified because of the narrow range of the survey around $\delta
= -1\degr$).  The RRLS in the clump cover a range of distances from
the Sun ($D$) with the greatest concentration at $D \sim 19$ kpc
\citep{vivas06}.  Using photometry from the SDSS, Newberg et al. (2002)
independently discovered an overdensity in F type main-sequence stars
in approximately the same direction and at a similar distance, which
they called S297+63-20.5.  They suggested that it may be related to
the small group of RRLS identified earlier by \citet{vivas01}, and
several authors have subsequently suggested that the 12.4 hr clump of
RRLS and S297+63-20.5 are the same substructure or at least
related. Also using the SDSS photometric catalogue, Juric et al (2005,
2008) \nocite{juric05old,juric05} later identified a very large
overdensity of main sequence stars toward Virgo (the {\it Virgo
Overdensity} (VOD) ).  On the sky, the VOD encompasses both the 12.4
hr clump and S297+63-20.5. But, according to \citet{juric05}, the VOD
is spread along the line of sight from $\sim 6$ to $\geq 20$ kpc,
which is a larger range than either the 12.4h clump or S297+63-20.5.
The spectroscopy of RRLS in the 12.4 hr clump by \citet{duffau06}
provided the first evidence of a moving group in the Virgo direction,
which they called the {\it Virgo Stellar Stream} (VSS).  The recent
spectroscopy of F type main-sequence stars in S297+63-20.5 by
\citet{newberg07} revealed a rich structure in velocity space, but
with a major peak that nearly coincides with the velocity of the VSS.
They tentatively conclude that S297+63-20.5, the VSS, and the VOD are
likely associated, if not the same. On the other hand, several authors
\citep{majewski03,martinez04,martinez07} have argued that the Virgo
substructures are parts of Sgr streams, while others
\citep{newberg02,newberg07,juric05,duffau06} have rejected this
hypothesis and suggested instead that they are due to a separate
merger event involving a dwarf galaxy.

We are carrying out a program to observe spectroscopically a large
number of QUEST RRLS in order to measure their radial velocities and
metallicities.  RRLS are arguably the halo tracer that can be most
precisely located in space, and with the addition of radial velocities
of even modest precision ($\pm 20$ \kms), they become powerful probes
of phase space.  This wonderful property of RRLS comes with a price, for
they are expensive observationally.  Complete light curves are
necessary to obtain both mean magnitudes (hence, precise distances)
and ephemerides. The later are necessary to isolate the motion of the
star in the sky from the pulsational velocity of the star by means of
the fitting of a radial velocity curve.  In this paper, we present
spectroscopic observations of bright RRLS ($V_0 < 16.1$, equivalent to
D $\lesssim 12.5$ kpc) in the part of the QUEST survey that overlaps
with the VOD.  After examining these data for halo substructure, we
discuss the relationships between our results and the previous
detections of substructures in Virgo and also consider the possibility
that they are related to streams from the Sgr dSph galaxy.  The RRLS
surveys by \citet{keller07} and by \citet{wilhelm07} provide important
data on parts of the VOD that are not covered by the QUEST survey.

\section{THE DATA}

\subsection{The Sample of RR Lyrae Stars}

According to \citet{juric05}, the VOD occupies $\sim 1000$ sq. degrees
of the sky, and its rough limits are shown in
Figure~\ref{fig-zone}. Notice that the SDSS does not extend south of
$b\sim58\degr$ at $l\sim 300\degr$ \citep[see Figure 6
in][]{juric05}. Thus, it is possible that the VOD extends farther in
that direction.  The region surveyed by QUEST is a $2\fdg 3$-wide band
centered at $\delta=-1\degr$. It passes over part of the VOD, as shown
in the Figure~\ref{fig-zone}. The dot-dashed line in
Figure~\ref{fig-zone} shows the southern limit of a second QUEST
catalog (Vivas et al. in preparation) which is centered at
$\delta=-3\degr$.  Eight stars from this second region are included in
our sample.

The region in common between the VOD and the QUEST survey
is roughly limited by galactic longitude in the range 
$260\degr \lesssim l \lesssim 340\degr$. At the 
declination of the QUEST survey,
those limits are equivalent to $169\fdg 5 < \alpha < 212\fdg3$. 
We have
therefore selected for spectroscopy all RRLS in the QUEST catalog 
\citep{vivas04} within these
limits in $\alpha$, and with magnitude $V\leq 16.1$. 
The constraint in magnitude was imposed by the
limitations of our major instrument, the SMARTS 1.5m telescope
(see \S~\ref{sec-observations}). The
magnitudes of this sample correspond to a range of distance 
from the Sun of $\sim 4$ to $12.5$ kpc
(using $M_V=0.55$, following Vivas \& Zinn 2006), which partly overlaps the 
range estimated for the VOD \citep[$\sim 6 - 20$ kpc,][]{juric05}.

Figure~\ref{fig-quest} shows the extinction corrected magnitude $V_0$
(equivalent to distance) of part of the QUEST RRLS in the Northern
Galactic Hemisphere, as a function of $\alpha$. The box encloses the
sample of 44 stars studied in this work. Within the box, all the stars
in the first QUEST catalog were observed spectroscopically, but only 8
RRLS (out of 38) from the second QUEST catalog were observed.  For
reference, we have indicated with solid symbols the location of the
RRLS that probably belong to the VSS \citep{duffau06}.  The stars in
our sample are located in the same area of the sky as the VSS, but are
closer to the Sun.  Most of them (the ones with $D>6$ kpc) are located
within the same region of the sky and distance range as the VOD.
Notice also in Figure~\ref{fig-quest} that part of the tidal stream of
the Sgr dSph galaxy lies in the background, at $\sim 50$ kpc.

\subsection{Spectroscopy \label{sec-observations}}

We obtained a total of 88 spectra of the 44 RRLS with 3 different
telescopes between 2001 and 2007. The instrumental setups of the
different observing runs are summarized in Table~\ref{tab-telescopes},
and the individual observations are detailed in
Table~\ref{tab-observations}.  Most of the spectra were obtained with
the SMARTS 1.5m telescope at Cerro Tololo Interamerican Observatory,
Chile (denoted as "SMARTS" in Table~\ref{tab-observations}).  We also
used observations obtained with the 1.5m telescope in La Silla, Chile
(denoted as "ESO") and with the Hydra multifiber spectrograph at WIYN
in Kitt Peak National Observatory, USA.  Most of the spectra were
taken at blue wavelengths, covering the Balmer lines (beginning in
H$\beta$) and the Ca II H and K lines.  However, we took advantage of
available time from other projects at the WIYN telescope and obtained
spectra using a setup which covered the red part of the spectra,
including the Ca triplet lines.  The WIYN observations are denoted in
Table~\ref{tab-observations} as "WIYN-B" and "WIYN-R" for blue and red
spectra respectively.

For each target star, we tried to obtain at least two spectra taken at
different phases during the pulsation cycle.  Thirty five out of the
44 RRLS have two or more spectra available.
Table~\ref{tab-observations} contains ID (same QUEST ID as in Vivas et
al. 2004), Julian date, exposure time, telescope, phase of observation
(see \S~\ref{sec-ephemerides}), heliocentric radial velocity with its
error (see \S~\ref{sec-rv}), and metallicity (see
\S~\ref{sec-feh}). The 8 stars from the unpublished second part of the
QUEST catalog have ID numbers higher than 700.

The spectrum of star \#254 indicates that is too cool to be a {\rrc}
star.  This is not surprising because a small fraction of the QUEST
\rrc stars are expected to be misidentifications of other types of
variable stars, mainly variable blue stragglers and W UMa eclipsing
binaries \citep[see][]{vivas04}. Star \# 254 was excluded from any
further analysis.

To avoid excessive broadening in the spectral features due to the
changing velocity of the star during the pulsational cycle, exposure
times were kept short, usually under 30 minutes. A few stars have
exposure times as long as 60 minutes, which is equivalent to
$\sim10\%$ of the pulsation cycle of a typical RRLS.  In all observing
runs, we observed several stars from the list of \citet{layden94}
which are both radial velocity and pseudo-equivalent width standards.

\subsection{Improving the Ephemerides \label{sec-ephemerides}}

Since good ephemerides are essential for fitting the radial velocity
curves, it is important to improve them as much as possible.  The
QUEST survey has a large number of epochs (an average 32 for our
targets), which should yield precise ephemerides.  In \citet{vivas04},
the time of maximum light (HJD$_0$) for the \rrc stars was set by the
observation with the brightest magnitude, which is not the best
procedure because it can be skewed by photometric error and by poor
sampling near maximum light.  In addition, for some \rrc stars, it was
necessary to list two possible periods because both produced
reasonable light curve shapes to the eye.

To obtain better light curve parameters, we have fitted a light curve
template to each \rrc star in our sample.  A sine curve produced a
poor fit because a typical \rrc stars has a phase difference between
maximum and minimum light of $\sim 0.4$.  Consequently, we constructed
a light curve template based on 9 of the best observed QUEST \rrc
stars, which did not suffer from period aliases or from a poorly
measured HJD$_0$.  This template was then fitted using $\chi^2$
minimization, while varying the amplitude, phase at maximum light,
magnitude of maximum light, and period around the values listed in the
QUEST catalogue.  We also explored the parameters around the $1/P \pm
1$ alias periods.

The ephemerides for the \rrab stars in the QUEST catalogue were more
secure because they were already determined from light curve fitting
and because period aliases were less of a problem for them.  However,
the period was not allowed to vary in the fitting.  This time it was
allowed to vary in the fitting, but the resulting light curves are not
significantly different from the old ones.  Only in 5 cases were the
changes in period equal to or slightly larger than $2 \times 10^{-5}$
days.  The ephemerides that yielded the best fits to the light curves
for the \rrc and \rrab stars have been adopted here and are listed in
Table ~\ref{tab-velocities}.

\subsection{Measurements of [Fe/H] \label{sec-feh}}

The metallicities of the RRLS have been measured from the spectrograms
following the methodology and calibration of \citet{layden94}, which
is based on Freeman \& Rodgers' (1975) \nocite{freeman75} modification
of the \citet{preston59} $\Delta$S method. The method, which involves
plotting the pseudo-equivalent width of the Ca II K line (W(K)),
corrected for interstellar absorption, against the mean
pseudo-equivalent widths of the $\beta$, $\gamma$, and $\delta$ Balmer
lines (W(H)), is described in detail in \citet{vivas05}.

In general, the type c RRLS are less numerous than the type ab in old
stellar populations, e.g. globular clusters, and because they are
harder to detect than type ab, their number in RRLS surveys of the
field suffer more from incompleteness.  \citet{layden94} did not
produce a [Fe/H] calibration for type c RRLS, which were ignored in
his study of the metallicities and kinematics of field RRLS.  There is
a sufficient number of type c stars in our sample that it is important
to measure [Fe/H] from our spectrograms.

The average type c variable is $\sim 600$K hotter and has a $\sim0.15$
dex larger surface gravity than the average type ab variable, although
the types overlap in both quantities, particularly when the type ab
variables are hottest during their pulsations.  The Layden calibration
may therefore require no modification when measuring type c RRLS.
This is supported by Kemper's (1982) \nocite{kemper82} demonstration
that the $\Delta$S method, which was devised for type ab variables, can be
applied to type c, and more directly by \citet{gratton04}, who found
that both types of variables in same globular cluster define a tight
sequence in a plot of K line strength against hydrogen line strength
with some intermingling of the types along the sequence.  However,
\citet{gratton04} defined their K line and H line strengths
differently than did \citet{layden94}.

To see if Layden's calibration can be applied to the type c, we
observed both type c and type ab variables in the globular cluster M3
with the Hydra multiobject spectrograph on the WIYN telescope.  While
the fiber setup targeted a large number of variables of both types in
this variable-rich cluster, the combination of partly cloudy weather
and variable fiber throughputs limited the number of usable spectra to
8 type c stars and 10 type ab.  Analyses of high dispersion
spectrograms of red giants \citep{sneden04} and RRLS
\citep{sandstrom01} and the tightness of the red giant branch in the
color-magnitude diagram of M3 indicate that there is very little if
any variation in [Fe/H] among the stars in this cluster.  For our
purposes, we assume that there is no real star to star scatter and
investigate whether Layden's calibration yields the same [Fe/H] when
applied separately to the two types of RRLS.

Our measurements for the M3 RRLS in the Layden pseudo-equivalent width
system are plotted in Figure~\ref{fig-feh}, where different symbols
are used to depict the type ab and type c stars.  The mean [Fe/H]
obtained from all 10 of the type ab variables is $-1.78$ with a
standard deviation of 0.27.  The large scatter is undoubtedly due to
the low S/N of some of the spectrograms.  The 5 type ab RRLS (filled
squares in Figure~\ref{fig-feh}) for which we have the highest S/N
spectrograms ($50 > S/N > 17$ at the CaII K line) yield
$\langle$[Fe/H]$\rangle = -1.72$ with a more reasonable std. dev. of
0.14.  The 8 type c variables yield $\langle$[Fe/H]$\rangle = -1.66$
(std. dev. $= 0.17$), and the 5 that have spectrograms with $41 > S/N
> 17$ yield $\langle$[Fe/H]$\rangle = -1.63$ (std. dev. $= 0.19$).  
The high S/N samples differ in $\langle$[Fe/H]$\rangle$ by $0.09 \pm 0.11$, 
which suggests that the offset, if any, between the type ab and c variables is small.
Since these mean values agree well with Zinn \& West's (1984)
\nocite{zinn84} result for M3 ([Fe/H] = -1.66 with an internal
precision of $\pm 0.06$), Layden's calibration appears to be equally
applicable to type c and type ab RRLS.

The metallicities that we have derived from the spectrograms of our
program stars are listed in the last column of
Table~\ref{tab-observations}.  The majority of the spectrograms have
$S/N > 20$, and from the experiences with the M3 variables, we
estimate that the precision obtained from one spectrogram is $\pm
0.15$ for the type ab variables and $\pm 0.20$ for the type c.  The
lower precision for the type c RRLS is consequence of the convergence
of the constant [Fe/H] lines at large W(H) (see Figure~\ref{fig-feh}).
A few spectrograms have $S/N \sim 15$, and the [Fe/H] values derived
from them are marked with colons in Table~\ref{tab-observations}.
Layden's calibration breaks down for type ab variables that are
observed on the rising branch, where the strength of the K line is
reduced by the relatively large effective gravity and shock waves may
produce emission in the cores of the H lines that reduce their
equivalent widths.  We have discarded measurements [Fe/H] that were
made at phases where these effects are likely to be important.  The
surface gravity variations of the type c variables are milder, and the
equivalent widths of the H lines are not affected by emission.  None
of their measurements have been discarded.

\subsection{Radial Velocities \label{sec-rv}}

Radial velocities for all blue spectra were obtained by Fourier
cross-correlation (IRAF’s {\sl fxcor} task) with radial velocity
standard stars of A and F spectral type, in the spectral range $3800 -
5200$ \AA.  The radial velocity standards were separated in two groups
according to their effective temperature.  Each target star was
cross-correlated with 6 to 14 different standard spectra of the group
with more similar effective temperature.  For each spectrum, we
obtained a mean radial velocity, $V_r$, by weighting the result from
each standard by the cross-correlation error returned by {\it fxcor}.
This error depends on both the S/N of the spectra and the similarity
between the target star and the standard.

In order to estimate the radial velocity errors produced by the
different telescope-spectrograph combinations, we cross-correlated
each spectrum of the radial velocity standards with the spectra of the
other standards, taking care to avoid cross-correlating two spectra of
the same star.  Reassuringly, the differences between the mean
velocities from these correlations and the literature values were
close to zero for all of the standards with each instrumental
setup. We adopted the standard deviations of these differences as the
minimum error in velocity that we can achieve with a particular setup
for our target stars (16, 8, and 12 km/s for the SMARTS, WIYN blue,
and ESO 1.5m spectra, respectively). Because these errors include the
uncertainties in the zero-points of the wavelength calibrations, we
believe that these values better reflect the true errors than the
means of the cross-correlation errors from {\it fxcor} , which are
smaller.

Cross-correlation was not used to calculate the radial velocities of
the red Hydra spectra because the subtraction of the strongest
night-sky lines left residuals in some wavelength regions.  Instead,
Gaussian line profiles were fitted to the unblended Paschen lines of
hydrogen and/or the two strongest lines of Ca II, which however at the
resolution of the spectrograms are each blended with a Paschen line.
The Ca II lines were only measured in spectra that were taken at the
coolest phases of the type ab variables when the Paschen lines are
much weaker than the Ca II lines.  Before computing the radial
velocities from the Ca II lines, we made slight adjustments to their
wavelengths to take into account the blending.  These were estimated
from high S/N spectrograms of bright stars of known velocity whose
spectra resemble the type ab variables at minimum light.  To test this
method, we obtained with the SMARTS 1.5m telescope red spectra of 6
bright type ab RRLS during cloudy weather.  The velocities obtained
from the measurements of the Ca II lines did not systematically
deviated from the ones measured from the Paschen lines alone, and the
systemic velocities (see below) obtained for these stars agreed with
values in the literature.  For the red Hydra spectra, the standard
deviation of the mean of the velocities given by the individual lines
ranged from 4 to 10 \kms, which is consistent with the results
obtained from similar spectra of bright stars of known velocity.

The best technique for obtaining the radial velocity of the center of
mass (the systemic velocity, $V_\gamma$) of a type ab RRLS is to
integrate the radial velocity curve given by weak metal lines over the
whole pulsational cycle, and the results obtained for RRLS in globular
clusters \citep[e.g.][]{storm92} are consistent with the radial
velocities of the clusters to within the internal velocity dispersions
of a few \kms. In our low resolution spectrograms, the weak lines
produce poorly determined peaks in the cross-correlations, whereas
well-defined ones are produced by the Balmer lines and other strong
features.  Furthermore, we could not obtain sufficiently large numbers
of spectra to determine the velocity curve for each star.  For the
\rrab, we adopted instead the method described by \citet{layden94},
which involves measuring the velocities at a few phases ($\phi$) from
low resolution spectrograms and then fitting them with the radial
velocity curve of the ab variable X Arietis that \citet{oke66} obtained
from measurements of $H\gamma$.  Observations made near maximum light
($\phi< 0.1$ and $\phi > 0.85$) were not used in these fits because
the radial velocity curve undergoes a large discontinuity in this
region \citep[see for example][]{smith95}.  The velocity curve that is
determined from the H lines has a larger amplitude and differs in
other ways from the one given by the metals lines 
\citep[see][]{oke62,oke66}.  Nonetheless, Layden's (1994) comparisons of his
measurements of $V_\gamma$ with literature values for the same stars
indicates that this method produces results that are consistent with
the errors of the measurements and the fits of the velocity curve.  In
the case of the template star X Ari, the systemic velocity is equal to
the H line velocity at $\phi = 0.5$, but there is some small star to
star variation in this value.  In our calculation of the errors in
$V_\gamma$, we have included terms to account for this variation and
for the variation in the amplitudes of the radial velocity curve
\citep[see][]{vivas05}.  For \rrc stars, which have much smaller
velocity amplitudes than the \rrab, there are at most small
offsets between the velocity curves given by the metal lines and the H
lines.  To measure $V_\gamma$, we fitted a template constructed by combining
the radial velocity curves of T Sex and DH Peg
\citep[see][]{duffau06}.

Figure~\ref{fig-lc} shows the best fit for each star having two or
more observations. These plots include all of the radial velocity
measurements for a star, even if they were not used in the fits
because they are close to $\phi=0$.  In each case, these points lie
within the expected range, which provides a check that our
measurements do not suffer from large systematic errors.  Notice also
that six of the stars were observed with two different telescopes, and
the consistency of the measurements, as seen in Figure~\ref{fig-lc},
is also an indication that large systematic errors are not present.

In the fitting of the radial velocity curves, the data were weighted
by their errors.  The mean difference between the data points and the
radial velocity curve, $\sigma_{\rm fit}$, in most cases was of the
same size, or smaller, than the typical errors in the individual
measurements (see Table~\ref{tab-velocities}).

For the \rrc, we calculated the error in the systemic velocity by
combining the individual $\sigma_r$ in the usual manner for
calculating the $\sigma$ of a weighted mean value.  For both types of
stars, if $\sigma_{\rm fit}$ was larger than $\sigma_\gamma$ we
adopted the former as our final error in the systemic velocity. The
mean error in the systemic velocities of all, but one, RRLS range from
7 to 28 \kms, with a mean value of 17 \kms. One star, the type c
\#174, had an unusually bad fit of the radial velocity curve and its
final error is 42 \kms.  Finally, we calculated the radial velocity
measured by an observer at the Sun who is at rest with the galactic
center, $V_{gsr}$ by
 
\begin{equation}
V_{gsr}=V_{\gamma}+10.0\,\cos{b}\,\cos{l}+225.2\,\cos{b}\,\sin{l}+7.2\,\sin{b};
\end{equation}

Our results are presented in Table~\ref{tab-velocities}, which lists:
ID, coordinates $\alpha$ and $\delta$, the extinction corrected mean V
magnitude, the type of RR Lyrae star, the period and Julian date at
maximum light (see \S~\ref{sec-ephemerides}), distance from the Sun
(D), number of spectra available and number of spectra actually used
in the fitting of the radial velocity curve, the error in the fit of
the radial velocity curve ($\sigma_{\rm fit}$), the systemic velocity
($V_\gamma$) and its error ($\sigma_\gamma$), and the velocity in the
galactic rest frame, ($V_{gsr}$).

\section{DISTRIBUTION OF RADIAL VELOCITIES: SUBSTRUCTURES IN VELOCITY SPACE}

As noted above, the VOD encompasses the entire range in $\alpha$ and
perhaps a larger range of distance than our sample of RRLS.  The
spatial distribution of our sample of RRLS (see the box in
Figure~\ref{fig-quest}) does not reveal any regions of exceptionally
high density that are strong candidates for halo substructure.  There
is a roughly uniform band of stars that crosses the region at
distances between 7 and 10 kpc and also a region of relatively high
density near $12$ kpc within $180\degr<\alpha<195\degr$.  This second
feature is the near side of the large over-density in RRLS that we
have referred to as the ``12.4 hr clump''
\citep{vivas02,zinn04,vivas06}, which at ~19 kpc contains the VSS
\citep{duffau06}.  Figure 12 in \citet{vivas06} shows the location of
the 12.4 hr clump in the first band of the QUEST survey, but only two
of the stars identified there as probable clump members are in our
sample.  The addition of RRLS from the 2nd QUEST band makes the clump
even more prominent, and there are 9 stars in our sample that now
appear to be outliers of the clump.  If the band of stars at $D \sim
8.5$ kpc and the 12.4 hr clump are real halo substructures, many of
the stars should have similar values of $V_{gsr}$.

Figure~\ref{fig-vgsr_histo} is a histogram of $V_{\rm gsr}$ of the
whole sample of RRLS.  The mean velocity is $+27$ ${\rm km~s}^{-1}$,
and the standard deviation is $135$ ${\rm km~s}^{-1}$.  These values
disagree with recent measurements of samples of BHB stars in the halo
(see below).  For a sample of 1170 BHB stars, \citet{sirko04b} found a
line-of-sight velocity dispersion, $\sigma_{\rm los}=101.6$ ${\rm
km~s}^{-1}$, and $\langle{V_{\rm gsr}}\rangle=0$ \kms.
\citet{brown08} found a similar $\sigma_{\rm los}$ in their sample of
BHB selected from the 2MASS survey, although they noted that the
velocity dispersion increased to $117$ {\kms } if they selected only
the most metal poor stars in their sample to avoid thick disk
contamination.  The Shapiro-Wilk test for normality
\citep{shapiro65}\footnote{we used the algorithm by
\citet{royston95}.} indicates that the $V_{gsr}$ distribution of our
sample does not deviate significantly from a normal distribution.  It
seems safe, therefore, to use the student t-test and the F-test, which
reject with $> 90\%$ confidence that our sample has the same mean
velocity and velocity dispersion as the BHB sample of
\citet{sirko04b}.  However, the velocity dispersion of our sample is
not significantly different from the value \citet{brown08} obtained
for their most metal-poor BHB stars.  While these differences may be
another sign that RRLS and BHB stars in the halo have different
kinematics \citep[see below and][]{kinman07}, the halo substructure in
our sample is definitely a major contributor to its relatively large
velocity dispersion and its positive mean velocity.

While not statistically different from a normal curve, the histogram
in Figure~\ref{fig-vgsr_histo} is clearly unexpected for a random
selection of halo objects, for it has a very prominent peak at
$V_{gsr} \sim +230$ ${\rm km~s}^{-1}$.  In total, there are eight
RRLS, or 19\% of our sample that have $V_{gsr} > +180$ ${\rm
km~s}^{-1}$, which is very unlikely to be random occurrence (1 in
10,000, if $\sigma_{\rm los}=101.6$ ${\rm km~s}^{-1}$
and$\langle{V_{\rm gsr}}\rangle=0$).  This excess of RRLS with large
positive $V_{gsr}$ is most likely due to a halo substructure.  Other
likely substructures, which have less extreme values of Vgsr, are
revealed when distance and velocity are used in concert.

In Figure~\ref{fig-bin}, we present histograms of $V_{gsr}$ for these
3 distance intervals, where we have selected dividing lines that avoid
putting stars on the boundaries. The histograms in
Figure~\ref{fig-bin} do not resemble the normal curves expected of
random selections of halo stars, and in each case the Shapiro-Wilk
test rejects normality with greater than 98\% confidence.  When
considering the $D < 6.5$ kpc interval, it is important to recall that
QUEST survey is probably incomplete at the bright magnitudes of these
relatively nearby stars because of saturation of one or more detectors
in the QUEST mosaic camera \citep[see][]{vivas04}.  Detector
saturation was not a problem for the stars in the more distant
intervals, and the QUEST survey is expected to be $\sim95\%$ and
$\sim70\%$ complete for the type ab and type c variables,
respectively.  For these intervals at least, the departures of the
$V_{grs}$ distributions from the expected halo curves are therefore
probably due the presence of halo substructure.

\subsection{Identification of Groups}

To identify possible groups of stars, we implemented with only a few
modifications the algorithm described by \citet{clewley06} (the
Stellar Pair Search). In this technique, the distances of each star
from all others in the sample are calculated from the cartesian
Galactic coordinates $x, y$ and $z$. Pairs of
stars are ones that are separated by less than a critical distance,
$d_{crit}$ and have a difference in radial velociticy less than a
critical velocity, $v_{crit}$.  Pairs having one star in common form a
group. A group may be formed by many connected pairs, and the minimum
number of stars in a group, $N$, is 3 (two pairs connected by one
star).

To study the statistical significance of the groups we performed Monte
Carlo simulations of random samples of stars having the radial
distribution of the RRLS in the halo.  Specifically, we used the
density radial profile described in \citet{vivas06} for the direction
$l=300\degr$, $b=60\degr$, which has a slope of $-3.1$ and takes into
account the flattening of the inner halo.  We assigned to each
simulated star a random $\alpha$ and $\delta$ within the same limits
of our sample, and a random $V_{\rm gsr}$ drawn from a Gaussian
distribution with $\sigma_{\rm los}=101.6$ ${\rm km~s}^{-1}$ and
$\langle{V_{\rm gsr}}\rangle=0$. Each simulated sample was introduced
in the group detection algorithm described above. The experiment was
repeated many times, and we counted how many times the simulated
samples produced groups with characteristics similar to the groups in
the real sample of RRLS.

The choices for the parameters $d_{crit}$ and $v_{crit}$ require some
discussion. As we described above, the VOD covers the whole range of
right ascension of the sample studied here, which is $\sim
40\degr$. Any substructure in our data may be as large as this. At 7
kpc from the Sun, $40\degr$ is equivalent to 5 kpc, and at 12 kpc from
the Sun, it is 8 kpc. On the other hand, the uncertainty in distance
of the RRLS is $\sim 7\%$ \citep{vivas04}, which suggests a lower
limit of $d_{crit}\lesssim 0.8$ kpc.  The streams from either
destroyed globular clusters or dwarf galaxies are expected to be cold
substructures, with velocity dispersion $\lesssim 30$ \kms. Given the
errors in $V_\gamma$, $v_{crit}$ should be larger than 17 \kms.  Since
halo substructures are unlikely to have uniform sizes or velocity
dispersions, we ran the group detection algorithm and the Monte Carlo
simulations using combinations of $d_{crit}$ from 1 to 4 kpc in steps
of 0.5 kpc and $v_{crit}$ from 20 to 30 {\kms } in steps of 5 \kms.  A
group in the RRLS data was only considered real when less than 5\% of
the 5000 Monte Carlo simulations (a $2\sigma$ detection) produced a
group with $\geq N$ members within $\pm 20$ {\kms } of $|V_{gsr}|$, the
absolute value of the mean velocity of the group of real RRLS.  The
simulations therefore counted random groups near the $V_{gsr}$ of the
suspected group in the real data and ones near $-V_{gsr}$.  If a
group was detected with different sets of parameters, as happened in
most cases, we report here the parameters that gave the lowest
probability of random occurrence and/or the maximim number of group
members.  We restricted our sample to the 33 stars that have
$D\geq6.5$ kpc since the incompleteness that is caused by the
different saturation levels of the CCDs of the QUEST camera
\citep{vivas04} may effect the interpretation of results at closer
distances.

Two significant groups were found among our RRLS
(Figure~\ref{fig-groups}).  As we suspected from the histograms in
Figures~\ref{fig-vgsr_histo} and~\ref{fig-bin}, there is a group with
large positive velocity in our data. The maximum number of members
(8 RRLS) was obtained with $d_{crit}=3.5$ kpc and $v_{crit}=25$
\kms. Not one of the 5000 Monte Carlo simulations produced a similar
group.  The stars making up this group are spread in $D$ from $\sim
7.5$ to $\sim 12.5$ kpc (see Figure~\ref{fig-groups}).  Four of the
more distant stars in this group ($D>11$ kpc) are part of the near
side of the ``$12.4$ hr clump''.  The unusually high velocities of
these stars and that the fact that some of them are probably part of
the ``$12.4$ hr clump'', unquestionably an over-density in RRLS,
suggest that they constitute a halo substructure that may be part of
the VOD (see \S~\ref{sec-dis}).  The combination of large $V_{gsr}$
and a spread in $D$ is consistent with a stellar stream that obliquely
cuts the line-of-sight \citep[see Fig. 15 in][]{harding01}.

The other significant group contains 9 RRLS.  While it is detected
with several choices for $d_{crit}$ and $v_{crit}$, the values of 1.5
kpc and 25 \kms, respectively, yield the lowest probability of a
random occurrence (only 0.7\% of the Monte Carlo simulations produced
similar groups, a $2.7\sigma$ detection). The relatively small value
of $d_{crit}$ of this group indicates that it is more concentrated in
space than the high velocity group (see Figure~\ref{fig-groups}).  It
lies $\sim 8.5$ kpc from the Sun and has a mean $V_{gsr}$ of $-56$
\kms. This group may be due to a stellar stream that cuts the
line-of-sight at nearly a right angle, which would explain the narrow
distribution in $D$ and the relatively low $V_{gsr}$ of the stars.

\subsection{Comparison with Blue Horizontal Branch Stars}

If any of the features described above are real substructures, then they should 
also appear in the 
distributions of the other types of stars that make up very old stellar populations.  
While BHB stars and RRLS are, of course, samples of horizontal branch stars that 
differ primarily in 
temperature, there are good reasons to suspect that they may not always trace the same stellar 
populations.  
The majority of the globular clusters in the outer halo have red HB’s, while the 
metal-poor globular clusters in the inner halo have primarily blue HB’s (the dependence of the second parameter of HB morphology on galactocentric distance).  Consequently, there are large 
differences in the frequencies of RRLS and BHB stars between the clusters in the inner and outer halos \citep{suntzeff91}.  The majority of the dwarf spheroidal 
satellite galaxies of the Milky Way and M31 have red HB’s and high frequencies of RRLS and low frequencies of BHB stars \citep{vivas06}.  \cite{kinman07} have suggested that this variation in the RRLS/BHB star ratio and the kinematic disconnect between the inner and outer halos 
(see \S~\ref{sec-intro}) may be responsible for the kinematic differences that they 
found between samples of RRLS and BHB in 
the same region of the Galactic halo.  These observations and the trend in RRLS/BHB suggest that the RRLS may be a denser tracer of substructure in the outer halo than BHB stars.  
Since a small fraction of outer halo globular clusters and dSph 
galaxies have blue HB’s, each with minority populations of RRLS, and since a small population of BHB stars exists in the red HB clusters, some correlation in phase space is expected between these types of stars in the outer halo.  On the other hand, an exact correspondence would be surprising.

The sample of BHB stars considered here was drawn primarily from the survey by \citet{sirko04a} that is 
based on photometry and spectroscopy from the SDSS.  This database of BHB stars is less complete than 
our RRLS sample because the SDSS spectroscopy targeted primarily quasars and galaxies.  The recent,
more complete sample of BHB stars studied by \citet{brown08} unfortunately
does not overlap with the Virgo region. From the \citet{sirko04a} catalog, we selected all BHB stars 
(35 in total) within the 
same range of $\alpha$ as the RRLS sample. In $\delta$ we expanded
the limits to $-3\fdg 5 < \delta < +4\fdg 0$. This is still a small range
compared with both the width of the region in $\alpha$, 
and the size of the VOD. We also constrained the distance 
range to $3.0 < D < 13.0$ kpc.  \citeauthor{sirko04a} estimate that their measurements have an 
average error in velocity of about $26$ ${\rm km~s}^{-1}$.

It has been established both from observations of globular clusters
and theoretical modelling that there is a relation between the
luminosity of BHB stars and their temperature in the sense that bluer
BHB stars are fainter than red BHB stars \citep[see for
example][]{brown05}.  When estimating the distances to their BHB
sample, \citet{sirko04a} adopted instead the same $L/L_\odot$ for all
BHB stars on the basis of some observational evidence that they
realized was in conflict with theoretical computations.
\citeauthor{sirko04a} noted that the adoption of the variation in
$L/L_\odot$ with effective temperature that is suggested by the
computations would lead to smaller distance moduli by an average of
$0.18$ mag.  \citet{kinman07} have used the color-magnitude diagrams
of globular clusters to derive a relationship between the $B-V$ colors
of BHB stars and their $M_V$ values, which is anchored on the same
$M_V$ for RRLS that we are using.  A comparison, after making the
necessary transformations between photometric systems, of the
\citet{kinman07} relationship to the one employed by \citet{sirko04a}
reveals that the difference in distance modulus is near zero at the
blue edge of the instability strip and again small at the highest
temperatures, but is $0.33$ mag (a factor of $1.16$ in distance) at
$g-r = -0.23$ or $B-V = 0$.  Because only a fraction of the sample of
BHB stars observed by \citet{sirko04a} {\it may} be effected by
distance errors, we have not attempted to recalibrate their distance
scale.  None of the following comparisons depend critically on an
exact match of the BHB and RRLS distance scales.

In a few of the fields that were observed with the Hydra multi-object
spectrograph on the WIYN telescope, we identified potential BHB stars
from the QUEST photometry.  The three stars that listed in
Table~\ref{tab-bhb} appear to be true BHB stars.  Their $(u-g)_0$ and
$(g-r)_0$ colors, as measured by the SDSS, are consistent with the low
gravities of BHB stars on the basis of Figure 10 in \citet{yanny00}.
>From our spectra, we measured the full widths of the $\beta$,
$\gamma$, and $\delta$ Balmer lines at 0.2 below the continuum level,
which are well-known diagnostics of surface gravity at the
temperatures of BHB stars \citep[eg.][]{wilhelm99}.  In addition to
the candidate BHB stars, we measured the WIYN spectra of two BHB stars
in the globular cluster M3 that are similar in $(B-V)_0$ color to the
candidates and of the bright A and F type stars that we employed as
radial velocity standards with the WIYN observations.  The candidate
BHB stars have smaller H-line widths than the main-sequence stars and
ones that are similar to the BHB stars in M3.  To permit direct
comparison with the \citet{sirko04a} sample of BHB stars, we adopted
their method to derive the distances to these BHB stars (see
Table~\ref{tab-bhb}).

In Figure~\ref{fig-bin_bhb}, we have plotted histograms of $V_{gsr}$
for the BHB stars from \citet{sirko04a} and Table~\ref{tab-bhb} that
lie in the same $D$ intervals as the ones in Figure~\ref{fig-bin}.  In
this diagram, there may be more smearing between the intervals than
for the RRLS because the absolute magnitudes of the BHB stars are more
uncertain (see above).  Neither of the two features identified in the
RRLS sample is recognizable in the BHB star data alone, which could be
due to the incompleteness of the sample or to low frequencies of BHB
stars in these groups. However, several BHB stars have $V_{gsr}$
values that coincide with the peaks in the RRLS histogram. Indeed,
when we introduced the combined sample of RRLS and BHB stars into the
group detection algorithm described above, the same two groups
identified in the RRLS sample alone appeared.  The group with negative
velocities now contains 14 members, 5 of which are BHB stars. The new
mean velocity is $-49$ \kms, which we adopt as the velocity of the
group. Only 5\% of the Monte Carlo simulations produced similar
groups.  The group with high positive velocities has 10 members (2 BHB
stars).  Similar groups were obtained in only 2 out of 5000
simulations (0.04\%).

The combined sample of RRLS and BHB stars also contains a third, small
but nonetheless significant, group having a mean radial velocity of
$-171$ \kms.  The group, which was detected with $d_{crit}=1$ kpc and
$v_{crit}=25$ \kms, contains only 3 members, all of which are BHB
stars located in the very narrow intervals of $D=11.1$ to $11.3$ kpc
and $\alpha=205\degr$ to $210\degr$. The Monte Carlo simulations found
simialr groups in $\sim4\%$ of the cases (a $2.5\sigma$ detection).
This group may be related to main sequence stars with similar
velocities (see below) and with 2 RRLS at essentially the same
distance and velocity but farther west ($\alpha\sim 187\degr$).  Since
this group lies near the faint limit of our sample, it may be the near
side of a larger substructure.

\subsection{Comparison with main-sequence stars \label{sec-ms}}

The recent investigation of the VOD region by \citet{newberg07} has
provided radial velocity measurements for many stars with the colors
of F type stars in two SEGUE fields, at ($l,b$) = ($288\degr,
62\degr$) and ($300\degr, 55\degr$), one of which overlaps with the
region we are considering here (see Figure~\ref{fig-zone}).
\citet{newberg07} make the very reasonable assumption that the vast
majority of these stars are probably near the main-sequence turnoff in
the color-magnitude diagram.  Because such stars have a significant
spread in absolute magnitude, their distances are much less precisely
known than either the RRLS or the BHB stars. The big advantage of
using these stars is that they are much more numerous than either RRLS
or BHB stars.  Below we examine their data for evidence for the major
groups described above.

\paragraph{Group at \boldmath $V_{gsr}=+215$ \kms.}

This group is probably the most remarkable feature seen in our data
since very few stars with such high velocities are expected in a
sample of halo stars (1 or 2 in our sample of RRLS).  The 10 RRLS and
BHB stars in this group have a mean velocity of $\langle{V_{\rm
gsr}}\rangle = +215$ \kms, and a standard deviation of $25$ \kms,
which is not much larger than the typical error in our radial
velocities ($17$ \kms).  As noted above, this group appears to be
spread along the line of sight, and it may be significant that in the
sample of RRLS studied by \citet{duffau06}, there are 3 stars with
$V_{gsr}$ values consistent with membership in this group that have
$D$ between 16 and 17 kpc.

In their discussion of the F-type stars, \citet{newberg07} briefly
mention an excess of high velocity stars, which we suspect is related
to this group of RRLS and BHB stars.  In their Figure 9,
\citet{newberg07} present for each of their fields histograms of
$V_{gsr}$ for the F stars in two ranges in $g_0$ magnitudes.  In the
histogram for the brighter group, which according to the distance
scale\footnote{these distance limits and all ones mentioned in
connection to the F-type main-sequence stars have substantial
uncertainties. The distance scale of \citet{newberg07} is based on the
luminosity function of Sgr tidal debris \citep{newberg02}, which
suggests that the assumption of one absolute magnitude for the F-type
stars yields a distance uncertainty of 16\%.  If however F-type stars
come from a more heterogeneous population, resembling for example the
globular clusters considered by \citet{bell07} that span a range in
[Fe/H], then the distance uncertainty may be as large as 40\% } of
\citet{newberg07} corresponds to 11 to 14 kpc, there are 8 F type
stars with $V_{gsr}$ consistent with assignment to our high velocity
group.  It is probably significant that all 8 of these stars are found
in the more northern of the two fields (see Figure~\ref{fig-zone}),
the one that overlaps with the region studied here (2 stars of even
higher $V_{gsr}$ are present in this field as well).  Since neither
one of the two F-star fields has any stars with $V_{gsr} < -200$ \kms,
there is an asymmetry toward large positive velocities in the same
region of the sky where we find one in the RRLS sample.  Among the
fainter F type stars, which lie at distances between 14 and 18 kpc,
there are only 3 stars in the 2 fields combined with $V_{gsr}$ within
the range of this group.

We have been able to determine metallicities for 6 of the 8 RRLS in
our sample that appear to be members of this group. They have a mean
$\langle {\rm [Fe/H]} \rangle = -1.55$, with a standard deviation of
0.15 dex, which is of the same size as the uncertainty in our [Fe/H]
measurements.  This narrow distribution is much more consistent with
the debris from a destroyed globular cluster than from a dwarf galaxy.
Two of the 3 high-velocity RRLS measured by \citet{duffau06} have
metallicities that are similar to the metallicities of the brighter
RRLS. The other one has a much lower ([Fe/H]$<-1.95$), which may be a
sign that it is unrelated.  We are observing more RRLS in the range
$12 < D < 16$ kpc, with the expectation that more velocity and
metallicity measurements will reveal the origin of this substructure.

\paragraph{Group at \boldmath $V_{gsr}=-49$ \kms.}

This is the most populated group that we find in our data, containing
9 RRLS and 5 BHB stars with velocities in the range $-80 < V_{gsr} <
-10$ \kms. The mean velocity of this group is $\langle{V_{\rm
gsr}}\rangle = -49$ \kms, with a standard deviation of $22$ \kms. The
contamination from unrelated halo stars is expected to be relatively
high at the low velocity of this group, and assuming the normal
distribution in Figure~\ref{fig-vgsr_histo} for such field stars, we
estimate that only about 50\% of the stars in this velocity interval
are probably members of a stream.  Thus, our estimates for both the
mean velocity and the velocity dispersion are likely to be skewed by
this contamination. The 9 RRLS in this group have $\langle {\rm
[Fe/H]} \rangle = -1.72$, with a standard deviation of 0.28 dex.
There is not a strong peak in the metallicity distribution that would
suggest that the large dispersion is caused by contamination by
non-members of an otherwise narrow distribution.

\citet{newberg07} found a significant peak at $-76$ {\kms } in the
velocities of the F-type stars, which considering the uncertainties is
close to the peak that we find.  While this suggests membership
in the same halo substructure, it is not altogether clear that they
overlap spatially.  This feature is most clearly evident in the
southern of the two fields investigated by \citet{newberg07}, the one
that is outside the region investigated here, and among the bluer and
brighter sample of F-type stars ($11\lesssim D \lesssim 14$ kpc if
they have $M_g = +4.2$).  On the other hand, the RRLS in this group
have $7.5 < D < 9.5$ kpc, which place them in the foreground of the
F-type stars, unless the F-type stars are less luminous than
suspected.  There is better spatial coincidence between the F-type
stars and a group of fainter RRLS that were measured by
\citet{duffau06}, which have similar $V_{gsr}$.  The relationships
between the F-type stars, the RRLS in \citeauthor{duffau06}, a sample
of M giants, and the tidal stream from the Sgr dwarf galaxy are
discussed at length by \citet{newberg07}, who suspect that this
feature may be a new halo substructure that is unrelated to Sgr.  The
feature that we have identified appears to be unrelated to Sgr as well
(see below).  More observations are required to clarify its
relationship to the feature described by \citet{newberg07}.

\paragraph{Group at \boldmath $V_{gsr}=-171$ \kms.}

The 3 BHB stars in the group have basically the same values of $D$,
11.2 kpc, but given the distance uncertainty a spread of $\sim 2$ kpc
cannot be excluded. \citet{newberg07} identify a peak in their F star
data at $V_{gsr}=-168$ \kms, and in this case, there are good reasons
to believe that the features are the same.  A peak is seen in both of
their fields, although it is more prominent in the southern field, the
one that is outside our region, than in the northern.  It is only seen
in the brighter F stars that have $0.2 < (g-r)_0 < 0.4$.  According to
the distance scale of \citet{newberg07} these stars lie at $D$ between
11 and 14 kpc, which agrees well with the distances of the BHB stars.

\subsubsection{Does the VSS extends to closer distances?}

The F star data show a very prominent peak at $130$ {\kms } in both
fields, and \citeauthor{newberg07} consider this to be the velocity of
S297+63-20.5 feature.  On the basis of the spatial coincidence between
S297+63-20.5 and the VSS and the near coincidence in $V_{gsr}$, they
conclude that these features are probably the same or physically
associated.  However, \citet{newberg07} note that the $30$ {\kms }
offset in $V_{gsr}$ between S297+63-20.5 and the VSS requires
explanation.

The peak at $130$ {\kms } is seen in the faintest and the reddest
group of F stars, which have $g_0$ between 20 and
20.5. \citet{newberg07} note, however, that the most of these stars
actually have $g_0 < 20.3$, which corresponds to $D < 16.7$ kpc on
their distance scale.  Since the core of the VSS is located at about
19 kpc \citep{duffau06}, it is possible that $30$ {\kms } offset is
due to a gradient in $V_{gsr}$ with $D$.  Some evidence for this is
provided by the data for the brighter and bluer F star data (see
Figure 10 in \citeauthor{newberg07}), which has a prominent peak at
$\sim 150$ \kms.  The histograms in Figures~\ref{fig-bin} and
~\ref{fig-bin_bhb} indicate there are 5 stars (2 RRLS and 3 BHB stars)
having a mean velocity of $+143$ {\kms } which, within uncertainties,
is identical to the peak seen in the bright F star data. All 5 RRLS
and BHB stars are located near the far distance limit of our sample at
$D >11.8$ kpc and within the near boundary of the "12.4 hr clump" of
RRLS. The 5 stars are indeed identified as a group by our algorithm
(with parameters $d_{crit}=2.5$ kpc, $v_{crit}=20$ \kms), but by
itself this group has only low significance because $\sim 25\%$ of the
Monte Carlo simulations produce similar groups.  While a chance
coincidence cannot be ruled out, the similarity in $\langle V_{gsr} \rangle$ between
this group and the peak in the bright F star data suggests that they
are related.  The group of RRLS and BHB stars provides a hint that the
VSS may extend to distances as short as $\sim12$ kpc.  A gradient in
$V_{gsr}$, with $V_{gsr}$ becoming less positive with increasing $D$,
is expected of a stellar stream that obliquely cuts the line of sight
\citep[see][]{harding01}.  It is also possible, however, that
observational error and sample selection may also explain the
$V_{gsr}$ offset. We are exploring the possible extension of the VSS
and the possibility of a gradient by measuring more QUEST RRLS at
distances between 13 to 18 kpc (Duffau et al. 2008, in preparation).

\section{THE RELATIONSHIP TO THE SGR STREAMS}

Soon after the first detections of over-densities of stars in the
direction of Virgo, several authors
\citep[eg.][]{majewski03,martinez04} speculated that they may be
debris from the disruption of the Sgr dSph galaxy.  Even then there
was some evidence to contrary, for \citet{newberg02} had noted that
the color of the turnoff in S297+63-20.5 was redder than the color of
the turnoff in the Sgr streams. \citet{duffau06} later argued that the
Sgr streams were not responsible for the VSS or the majority of the
other stars in their sample because models of the streams produced
very few stars at the observed positions and the predicted values of
$V_{gsr}$ did not match the ones observed.  The realization by
\citet{juric05} that the VOD has huge dimensions on the sky motivated
\citet{martinez07} to reexamine the Sgr connection.  Through detailed
modeling of the streams, they showed that ones that assume an oblate
shape for the gravitational potential of the Galaxy can explain the
VOD as the leading stream from Sgr as it plunges toward the Sun and
crosses the galactic plane in the solar neighborhood.  This model does
not, however, provide a good match to our observations, those of
\citet{duffau06}, and more generally the 12.4 hr clump of RRLS.

In the left-hand panels of Figure~\ref{fig-sgr}, we have plotted the
RRLS in the first band of the QUEST survey that lie within the
$\alpha$ range of our sample.  Closed symbols depict the RRLS that
have measured velocities from \citet{vivas05}, \citet{duffau06}, and
the present study.  These three datasets cover different range of
distances, $>40$, $16 - 20$ and $<12.5$ kpc respectively.  The small
crosses are the particles in the stream models of \citet{law05} under
the assumptions that the potential is oblate, spherical, or prolate.
The panels on the right compare the model predictions for $V_{gsr}$ as
a function of $D$ with the observed velocities.

The QUEST survey detects the Sgr leading arm at $D\sim 50$ kpc in the
range $195\degr < \alpha < 235\degr$ (see Figure~\ref{fig-quest} and
Vivas et al. 2005) and the 12.4 hr clump in the ranges $12 < D < 20$
kpc and $170\degr < \alpha < 200\degr$ \citep[see Fig. 12
in][]{vivas06}.  Note that there is not a blending together of these
features (see Figure~\ref{fig-quest}), as one would expect if they are
parts of the same stream, but instead they are separated by a region
of low density.  This is not a selection effect introduced by the
observing technique \citep[see][]{vivas04}.

If the leading arm crosses the galactic plane near the Sun, then the
stream stars at small $D$ will have large, negative values of
$V_{gsr}$ because they have fallen from $\sim 50$ kpc on orbits that
are nearly aligned with the line-of-sight from the Sun.  This
prediction of large negative velocity at small $D$ and the continuity
of the stream are two characteristics of this picture that should not
be model dependent.  \citet{martinez07} note that their models
reproduce the models of \citet{law05}, which because they are readily
available, we will use here to illustrate the salient points of this
picture.

Each of the \citeauthor{law05} models matches reasonably well the
positions and the velocities of the RRLS at $\sim 50$ kpc.  However,
both the oblate and the spherical models predict smoothly varying
densities of RRLS over wide ranges in $D$, which are in conflict with
the observed region of relatively low density of RRLS between 20 and
40 kpc \citep[see also][]{vivas05}.  These same models predict that
$V_{gsr}$ becomes more negative as $D$ decreases and reaches quite
large $\left |V_{gsr} \right |$ at $D < 20$ kpc.  Only one RRLS in our
sample and one in the sample observed by \citet{duffau06} have values
of $V_{gsr}$ that are consistent with these models.  Also, none of the
BHB stars in the region have highly negative velocities (see
Figure~\ref{fig-bin_bhb}). In the prolate model, the leading stream
crosses the galactic plane far from the Sun, which explains why it
predicts relatively few stars in the diagrams at $D < 20$ kpc.  We
conclude on the basis of these comparisons that none of the halo
substructures that we have detected are probably due to the leading
arm.  Moreover, based on the current Sgr models, our observations
imply that the shape of the Galactic potential can be neither oblate
nor spherical.

\citet{martinez07} note that in their oblate model the trailing stream
of Sgr crosses through the region of the VSS and coincides with the
VSS in $V_{gsr}$ ($\sim 90$ km/s).  This is a plausible explanation
for the VSS, if it is still viable after the oblate model is
reconciled with the observations presented here and with other recent
observations which indicate that the leading arm does not cross the
galactic plane in the solar neighborhood \citep{newberg07,seabroke08}.

\section{DISCUSSION \label{sec-dis}}

Kinematical information is crucial to detect substructures in regions
where no obvious spatial overdensities are present.  Taken together,
our RRLS sample, the BHB star sample of \citet{sirko04a}, and the F
star sample of \citet{newberg07} paint a consistent picture that
several separate halo substructures exist at $D \lesssim 13$ kpc in
the part of the VOD that is covered by the QUEST survey. The RRLS
sample of \citet{duffau06} and the same BHB and F star samples show
that some of these substructures extend to larger $D$. None of these
substructures seem to be related with the Sgr leading tail.

\begin{figure}
 \includegraphics[scale=0.81]{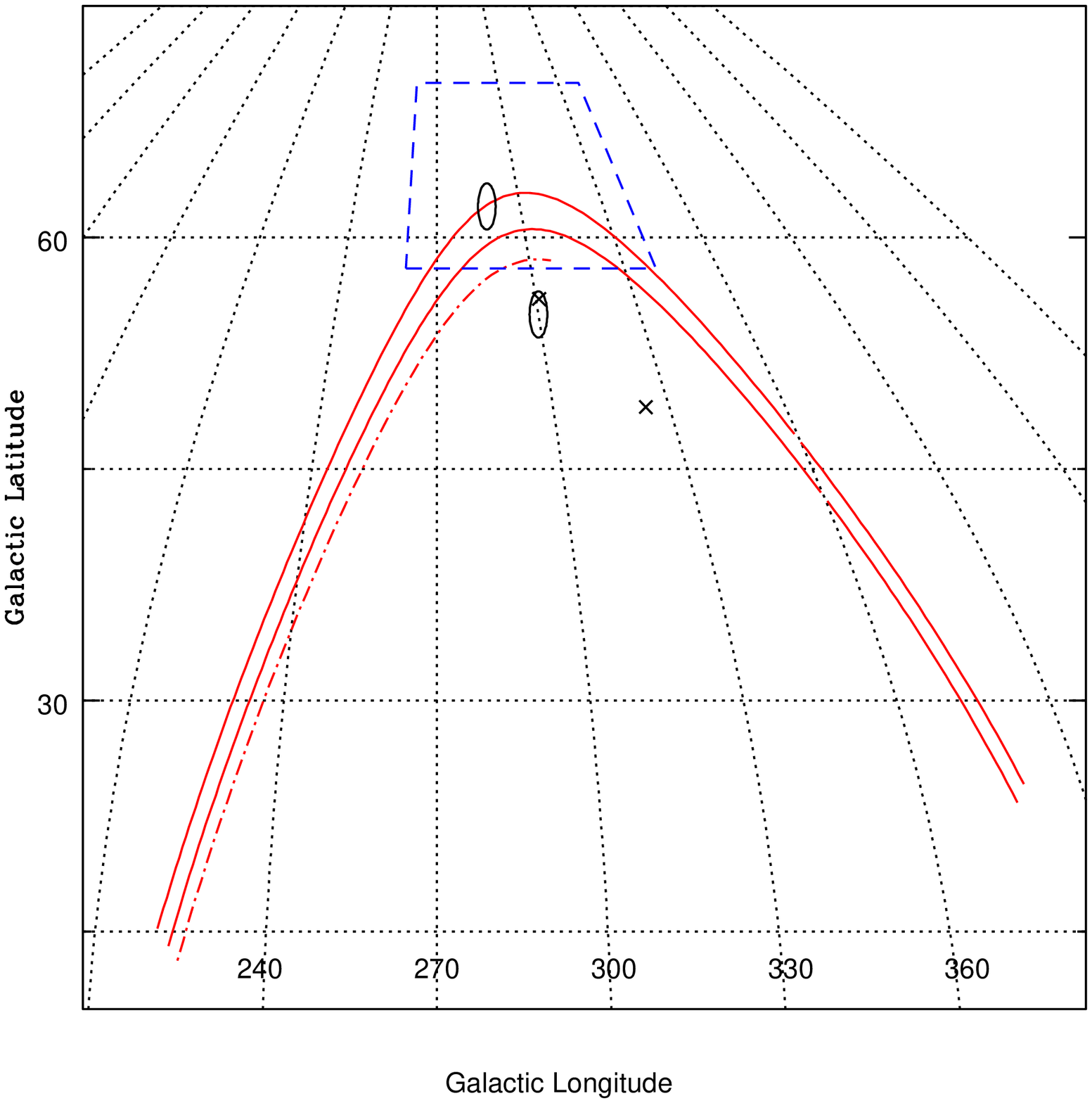}
 \caption{Map in Galactic coordinates showing the strip (solid lines) surveyed in 
the first QUEST
catalog in the Northern Galactic Hemisphere. 
The dot-dashed line indicates the region of the second (unpublished) QUEST catalog.
According to \citet{juric05}, the region of the VOD is 
roughly enclosed by the polygon.
The small ovals indicate the two regions where \citet{newberg07} looked for
radial velocity substructures among F stars ( $(l,b) = ($300\degr, 55\degr$)$ and $(288\degr,62\degr)$ ). The two $\times$ symbols indicate the locations where
\citet{keller07} found overdensities of RRLS at 16 and 19 kpc from the Sun.
}
  \label{fig-zone}
\end{figure}

\begin{figure}
 \includegraphics[scale=0.8]{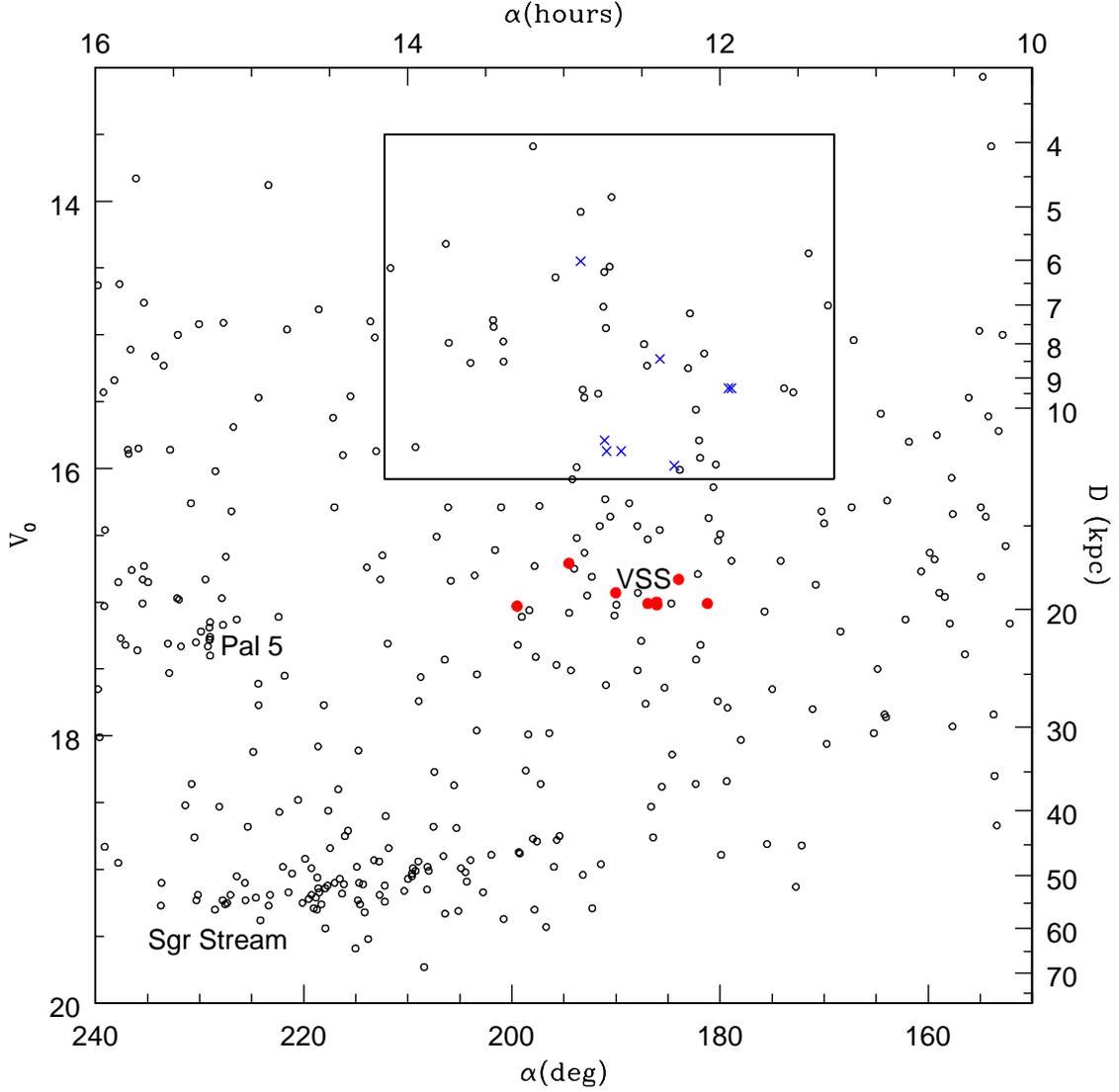}
 \caption{Spatial distribution of the QUEST RRLS in the range of right ascension 
from $150\degr$ to $240\degr$ \citep[from][]{vivas04}. 
The box encloses the bright
subsample ($V_0<16.1$) toward Virgo for which we obtained spectroscopy. 
The $\times$ symbols indicate the 8 RRLS from the second QUEST catalog with
available spectroscopy that are included in our analysis.
The large spatial substructures found among the fainter RRLS \citep[see][]{vivas06}
are indicated: the Sgr stream, the
globular cluster Pal 5 and its tails, and the ``$12\fh 4$ clump'', which contains the VSS.
We have indicated with solid circles the RRLS having velocities 
consistent with the VSS \citep{duffau06}.}
 \label{fig-quest}
\end{figure}

\begin{figure}
 \includegraphics[scale=0.68]{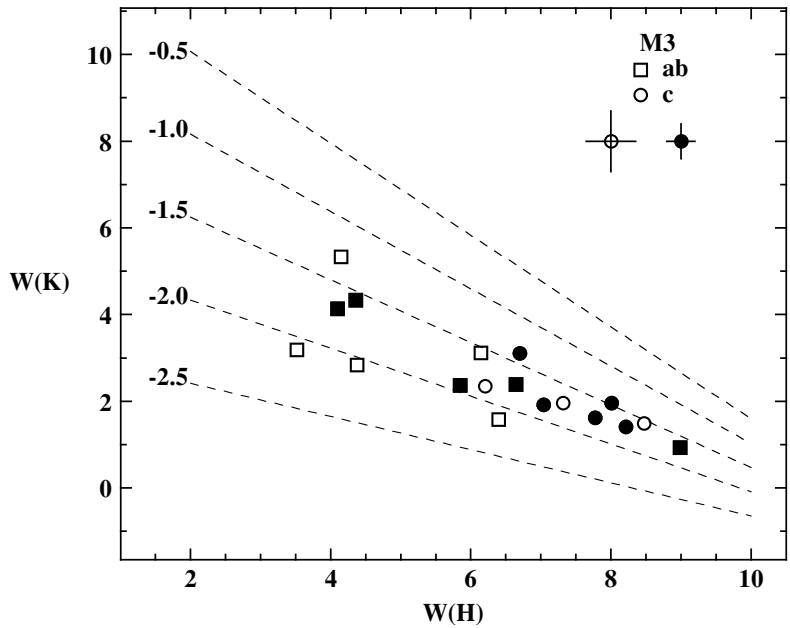}
 \caption{Calibration in the plane W(K) vs W(H) of the metallicity of RRLS, from
\citet{layden94}. Depending on their metallicity, 
\rrab stars lie along one of the dashed lines during the pulsation cycle (excluding
measurements during the rising branch). Symbols represent measurements of RRLS
variables in M3, with filled symbols corresponding to high S/N observations.
Typical error bars for the low and high S/N spectrograms are shown. 
This diagram shows that Layden's calibration is also valid for \rrc stars.}
  \label{fig-feh}
\end{figure}

\begin{figure}
 \plotone{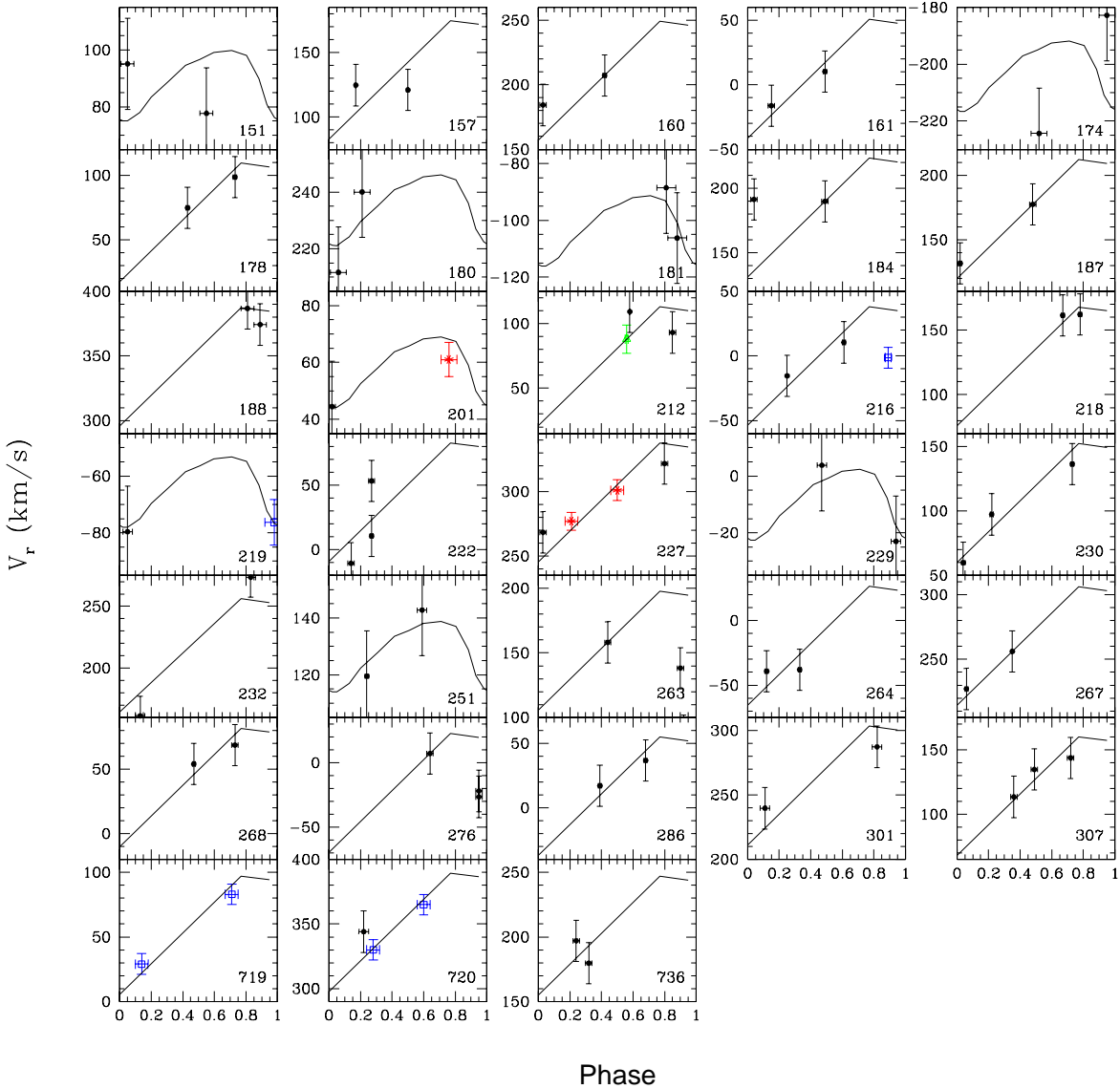}
 \caption{Radial velocity curves for the 33 stars having two or more spectroscopic observations.
The SMARTS observations are indicated with solid circles, the ESO observation is marked with an 
open triangle, the WIYN-R observations are asterisks, and the WIYN-B are open squares.  
The fitted radial velocity curves are indicated by solid lines.
All observations are shown including those
ones not used in the fitting of the radial velocity curve because they were taken near the phase of
maximum light (see text). Although they were not used in the fitting, they lie within the expected range of 
velocity.}
 \label{fig-lc}
\end{figure}

\begin{figure}
\plotone{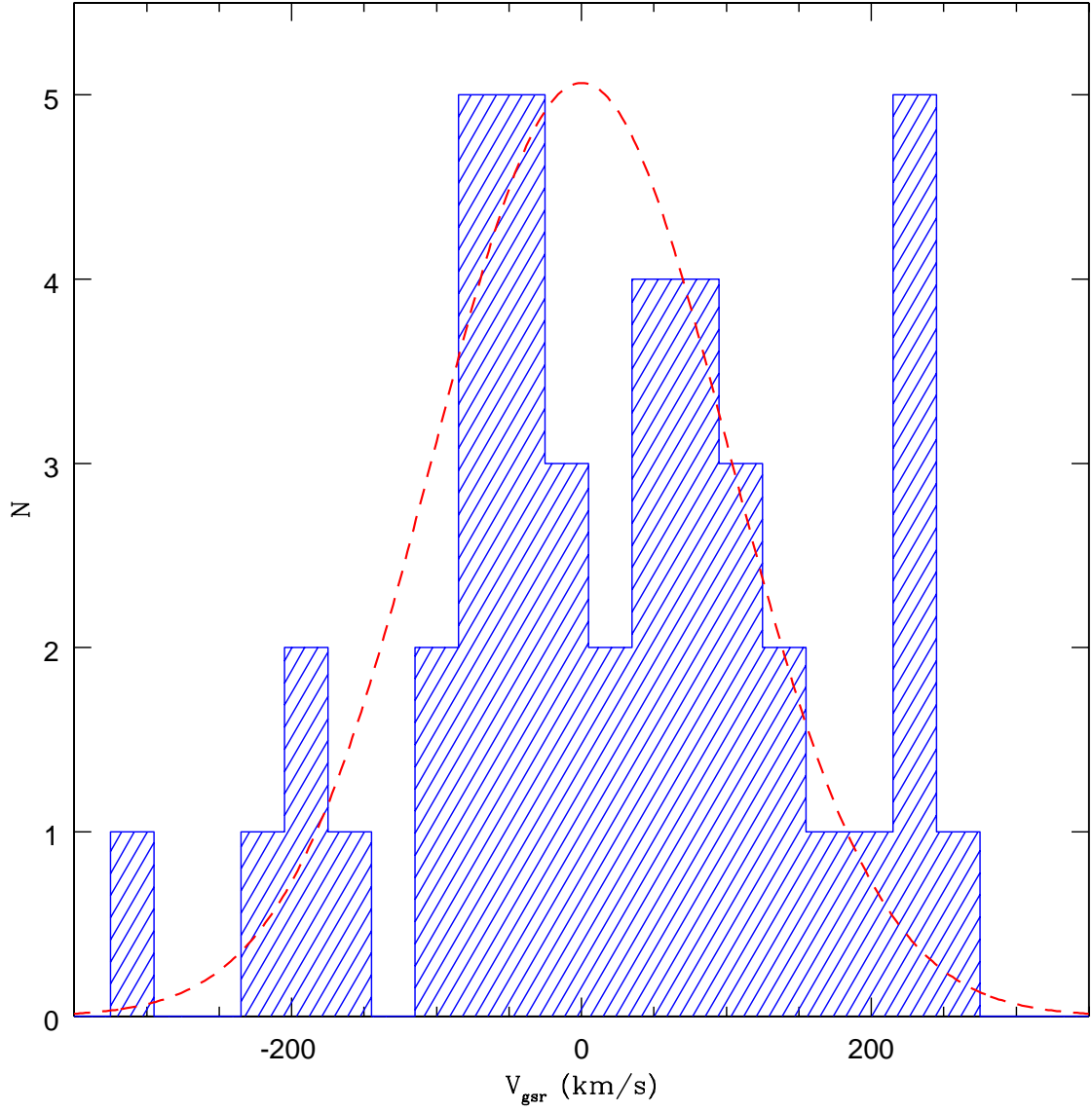}
\caption{Histogram of $V_{gsr}$ of the 43 RRLS. For reference,
the dashed line indicates the expected distribution of velocities in a random halo sample 
($\sigma_{\rm halo}=101.6$ ${\rm km~s}^{-1}$). The size of each bin is 30 \kms.}
\label{fig-vgsr_histo}
\end{figure}

\begin{figure}
 \plotone{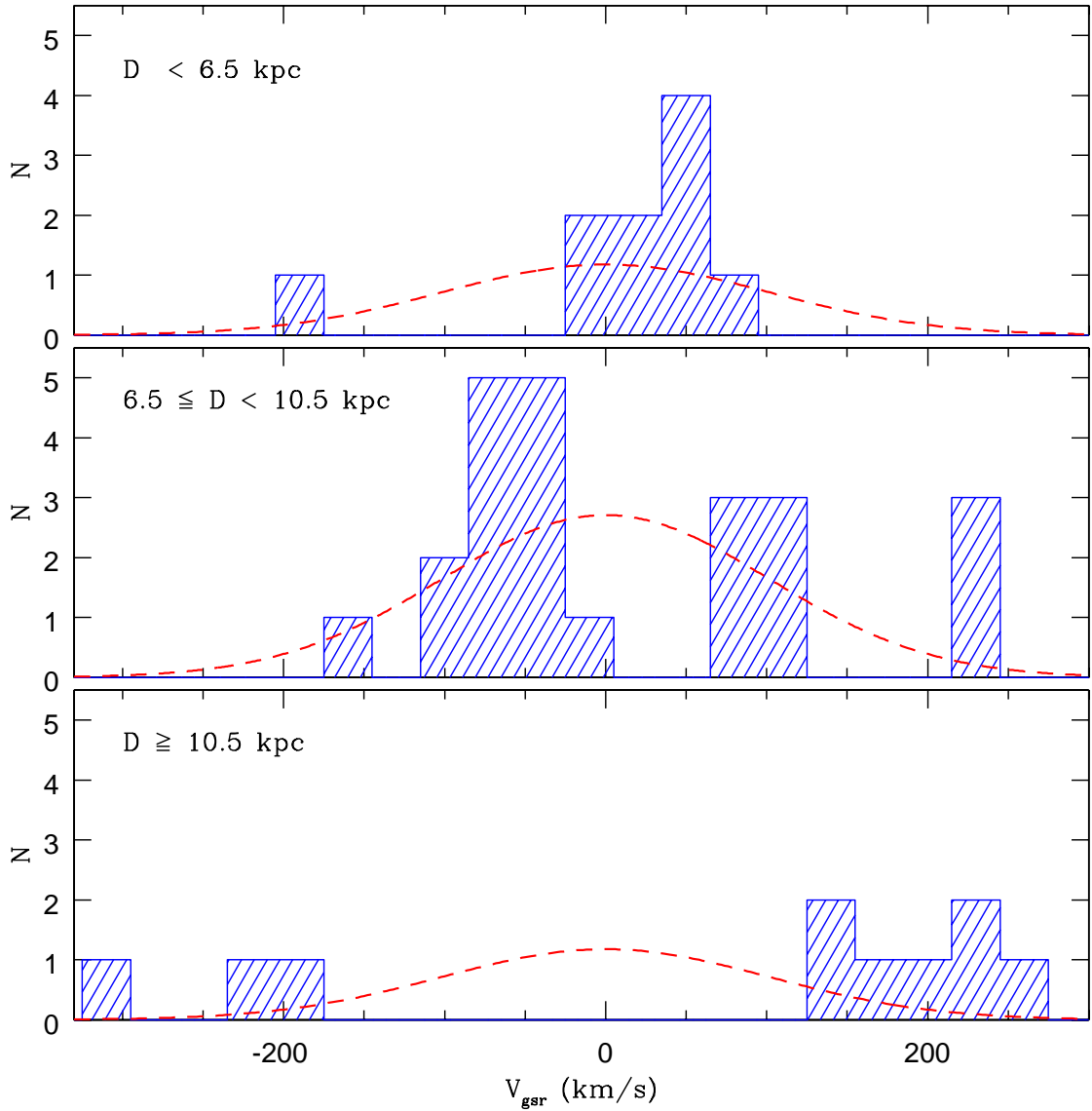}
 \caption{Histogram of $V_{gsr}$ of the RRLS in 3 distance bins. The dashed line indicates 
the expected distribution of velocities in the halo, normalized to the number of stars in each
range of distance.}
 \label{fig-bin}
\end{figure}

\begin{figure}
 \includegraphics[scale=0.66]{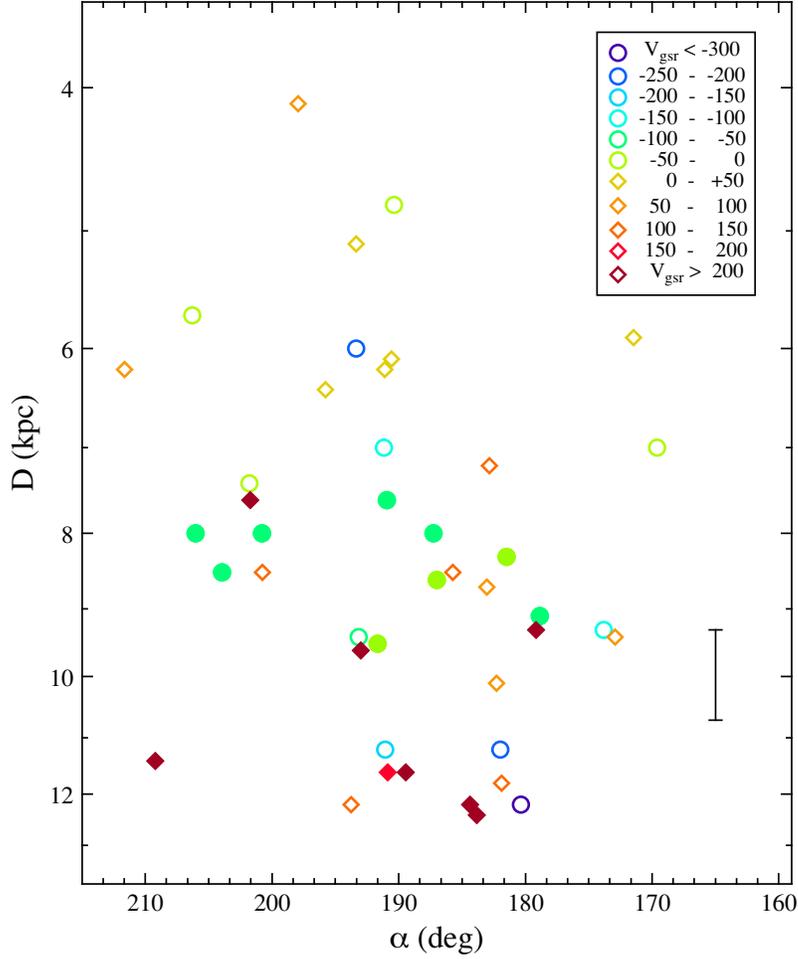}
 \caption{Distribution of distance from the Sun of the sample of RRLS 
as a function of right ascension. 
Circles and diamond symbols represent stars with negative and positive radial velocities respectively. Colors indicate stars with similar velocities in 
bins of 50 \kms. The two significant groups found in our data are indicated by
solid symbols, one of them have stars with very high positive velocities and the other
one has stars with velocities around $-60$ \kms.}.
\label{fig-groups}
\end{figure}

\begin{figure}
 \plotone{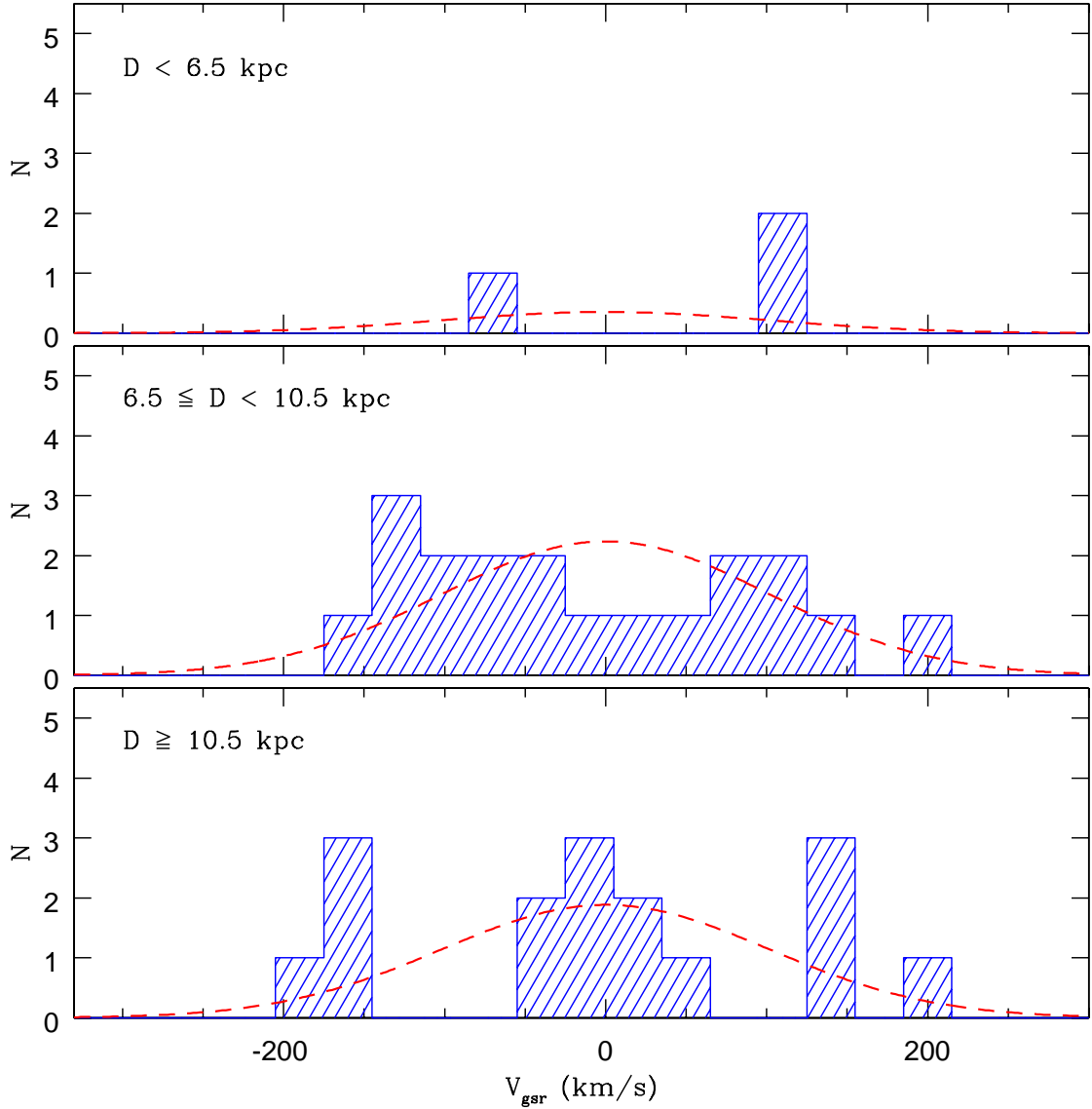}
 \caption{Histogram of $V_{gsr}$ of the BHB stars in the same 3 distance bins as Figure~\ref{fig-bin}.
The dashed line indicates 
the expected distribution of velocities in the halo, normalized to the number of stars in each
range of distance.}
 \label{fig-bin_bhb}
\end{figure}

\begin{figure}
 \includegraphics[scale=0.61]{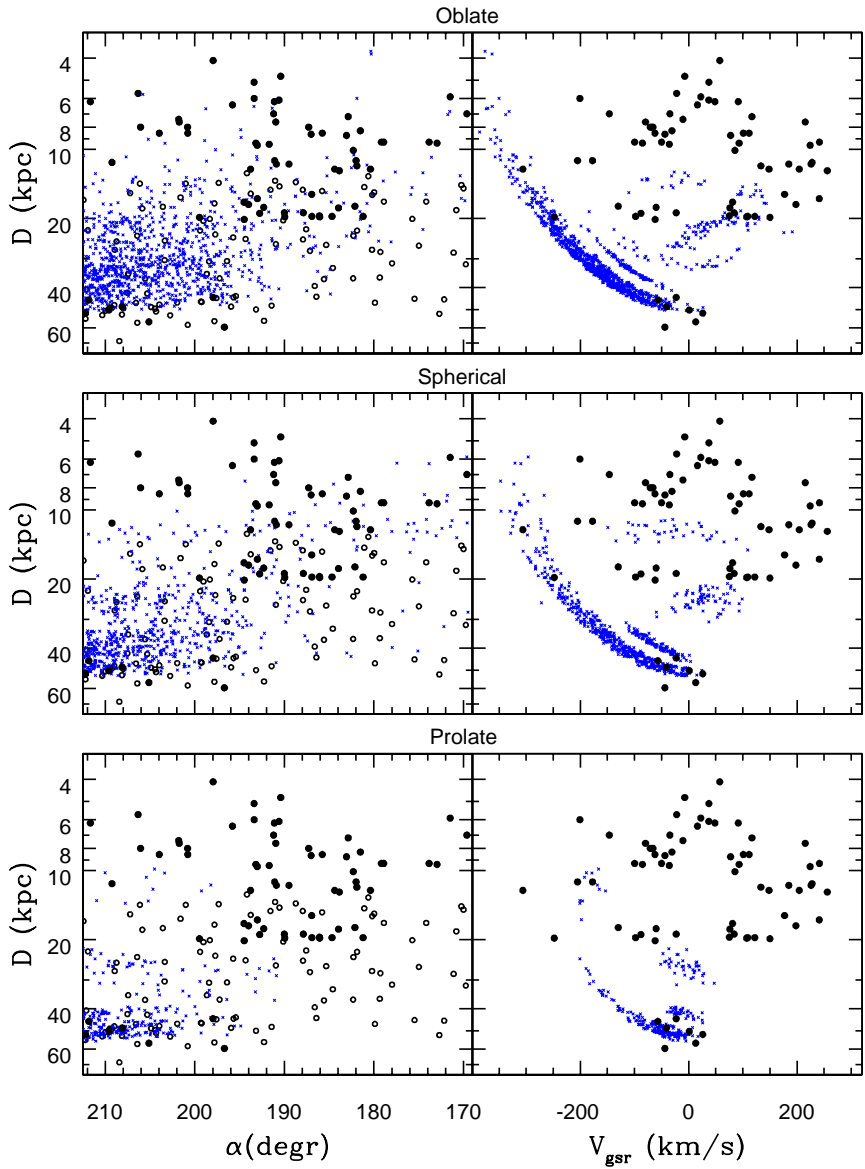}
 \caption{Comparison of the location and velocity of the QUEST RRLS with models of the Sgr
streams (small crosses) by \citet{law05},
which assume different shapes for the galactic potential
(top: oblate; middle: spherical, bottom: prolate). 
Filled symbols correspond to different sets of spectroscopic observations of
QUEST RRLS: stars with $D<12.5$ kpc (Table 3); $16<D<20$ kpc \citep{duffau06}, 
$D>40$ kpc \citep{vivas05}. The open circles in the left panels show the position of
QUEST RRLS that have not been observed spectroscopically.}
 \label{fig-sgr}
\end{figure}

The analysis of the SDSS photometric data by
\citet{newberg02,newberg07} make a compelling case that the region of
highest stellar density in S297+63-20.5 lies at $D \sim 18$ kpc.  This
is identical to within the errors of the region of highest density in
the 12.4 hr clump of QUEST RRLS, which is where \citet{duffau06}
discovered the VSS in velocity space.  The near coincidence in
$\langle V_{gsr} \rangle$ between the VSS and the largest peak in
$V_{gsr}$ that \citet{newberg07} observed in S297+63-20.5 provides
strong evidence that they are the same.  As noted by
\citet{newberg07}, the small offset ($\sim 30$ \kms) could be due to
measurement error and/or by the small sizes of the samples of
spectroscopically observed stars.  It is also possible that it is due
to a gradient in $V_{gsr}$ with $D$ (see above).  \citet{newberg07}
found the same velocity peak in their two spectroscopic fields, which
indicates that the S297+63-20.5/ VSS has considerable angular extent
on the sky.  Other evidence for this is provided by the recent RRLS
surveys by \citet{keller07} and \citet{wilhelm07}. \citet{keller07}
found 2 clumps of RRLS south of the VOD region (see
Figure~\ref{fig-zone}) at coordinates $(l,b) = (301\degr,56\degr)$ and
$(324\degr,49\degr)$ and distances 16 and 19 kpc
respectively. Velocity information is not yet available for these
clumps.  \citet{wilhelm07} identified an excess of candidate RRLS
having $V_{gsr} \sim 100$ {\kms } in the North Galactic Cap
($b>70\degr$). The results presented here indicate that the
S297+63-20.5/VSS feature may also extend along the line of sight to
distances as close as $\sim 11.8$ kpc.

We conclude that the VOD contains at least one relatively large
structure, S297+63-20.5 /VSS, which is intersperced with smaller halo
substructures that are unrelated to S297+63-20.5/VSS or to each other.
While these smaller structures might be peculiar to the VOD region, it
is more likely that they are common in the outer halo.  They are
better documented in the VOD direction simply because the presence of
S297+63-20.5/ VSS has drawn attention to this region.  In the case of
our sample of RRLS, the spatial distribution of the stars provided
only very marginal evidence of any substructure \citep{vivas06}.  The
picture changed dramatically when we used spatial and velocity data
together.  There is already strong evidence from the spatial
distribution of stars in the SDSS that much, if not all, of the outer
halo was accreted from the destroyed of satellite galaxies
\citep{bell07}.  Our investigation of the VOD region suggests that
the addition of velocity data will make the case even stronger.

\acknowledgments

This research was supported by the National Science Foundation under grants AST 00-98428 and AST 05-07364.  The availability of the SMARTS 1.5m telescope and its service observing mode made this project feasible.  We thank Charles Bailyn for his efforts to setup the SMARTS consortium and Fred Walter and Suzanne Tourtellotte for their assistance with SMARTS operations and data handling.  We are indebted to the service mode observers Sergio Gonzalez, Alberto Pasten, Claudio
Aguilera and Alberto Miranda for their fine operation of the telescope and the spectrograph.
We also thank the anonymous referee for helpful suggestions
to the manuscript.

\clearpage
\begin{deluxetable}{lccccccl}
\rotate
\tabletypesize{\small}
\tablecolumns{8}
\tablewidth{0pc}
\tablecaption{Telescopes and Instruments}
\tablehead{
\colhead{Observatory} & \colhead{Telescope} & \colhead{Instrument} & 
\colhead{Grating} & \colhead{Spectral Range} & \colhead{Resolution} & 
\colhead{\# Spectra/} & \colhead{Dates} \\
\colhead{} & \colhead{} & \colhead{} & \colhead{(mm$^{-1}$)} & \colhead{(\AA)} &
\colhead{(\AA)} & \colhead{\# Stars} & \colhead{}
}
\startdata
ESO/La Silla & 1.5m    & B\&C   & 600 & 3300- 5500 & 3.1 &   1/1  & 2001 Jun 17   \\
KPNO & WIYN        & Hydra  & 600 & 7100-10000 & 2.9 &   5/3  & 2003 Mar 11   \\
CTIO & SMARTS 1.5m & R-C    & 600 & 3532-5300  & 4.3 & 42/28  & 2003 May 1-9  \\
CTIO & SMARTS 1.5m & R-C    & 600 & 3532-5300  & 4.3 & 24/14  & 2004 Feb-Jun\tablenotemark{a}   \\
CTIO & SMARTS 1.5m & R-C    & 600 & 3532-5300  & 4.3 &   5/4  & 2006 Jun\tablenotemark{a}   \\
CTIO & SMARTS 1.5m & R-C    & 600 & 3532-5300  & 4.3 &   4/2  & 2007 Feb-Apr\tablenotemark{a}   \\
KPNO & WIYN        & Hydra  & 400 & 3500-6200  & 7.1 &   9/5  & 2007 Feb 20-21   \\
KPNO & WIYN        & Hydra  & 600 & 7100-10000 & 2.9 &   3/3  & 2007 Apr 27, May 2 \\
\enddata
\tablenotetext{a}{Service mode run}
\label{tab-telescopes}
\end{deluxetable}

\begin{deluxetable}{cccccrrc}
\tabletypesize{\small}
\tablecolumns{8}
\tablewidth{0pc}
\tablecaption{Individual Observations of the sample of RRLS\tablenotemark{a}}
\tablehead{
\colhead{ID} & \colhead{JD} & \colhead{Texp} & \colhead{Telescope} &
\colhead{Phase} & \colhead{$V_r$} & \colhead{$\sigma_r$} & \colhead{[Fe/H]} \\
\colhead{} & \colhead {(d)} & \colhead{(s)} & \colhead{} &
\colhead{} & \colhead{(${\rm km~s}^{-1}$)} & \colhead{(${\rm km~s}^{-1}$)} & \colhead{}
}
\startdata
201 & 2452711.7580 & 3600 &   WIYN-R & 0.76 &   58 &  5 &  \nodata \\ 
201 & 2452765.6208 & 1200 &   SMARTS & 0.02 &   44 & 16 & -1.95 \\ 
203 & 2452765.6385 & 1200 &   SMARTS & 0.02 &   13 & 16 & -1.40 \\ 
212 & 2452078.4665 &  900 &    ESO & 0.56 &   88 & 11 & -2.11 \\ 
212 & 2452760.6437 &  300 &   SMARTS & 0.58 &  109 & 16 & -1.99 \\ 
\enddata
\tablenotetext{a}{The complete version of this table is in the electronic
edition of the Journal.  The printed edition contains only a sample.}
\tablenotetext{:}{Uncertain measurements due to low SNR of the spectrogram. This measurement
was not taken into account for calculation of the mean [Fe/H] unless it was the only
available measurement for the star.}
\label{tab-observations}
\end{deluxetable}

\begin{deluxetable}{ccccccccccccccc}
\rotate
\tabletypesize{\scriptsize}
\tablecolumns{15}
\tablewidth{0pc}
\tablecaption{Position, Distance, Velocity and Metallicity of the 43 RRLS\tablenotemark{a}}
\tablehead{
\colhead{ID} & \colhead{$\alpha$} & \colhead{$\delta$} & \colhead{V$_0$} &
\colhead{Type} & \colhead{Period} & \colhead{HJD$_0$} & \colhead{$r_\odot$} &
\colhead{N} & \colhead{N$_{\rm fit}$} & \colhead{$V_\gamma$} & \colhead{$\sigma_{\rm fit}$} &
\colhead{$\sigma_\gamma$} & \colhead{$V_{gsr}$} & \colhead{[Fe/H]} \\
\colhead{} & \colhead {(2000.0)} & \colhead{(2000.0)} & \colhead{} &
\colhead{} & \colhead{(d)} & \colhead{(d)} & \colhead{(kpc)} &
\colhead{} & \colhead{} & \colhead{(${\rm km~s}^{-1}$)} &
\colhead{(${\rm km~s}^{-1}$)} & \colhead{(${\rm km~s}^{-1}$)} & \colhead{(${\rm km~s}^{-1}$)} & \colhead{}
}
\startdata
151 & 169.625160 & -0.764498 & 14.78 &  c & 0.29370 & 2451611.7002 &  7.0 & 2 & 2 &   90 & 28 & 28 &  -35 & -1.32 \\
157 & 171.487530 & -0.161508 & 14.39 & ab & 0.70840 & 2451571.8154 &  5.9 & 2 & 2 &  142 & 22 & 22 &   22 & -1.67 \\
160 & 172.957305 & -2.240574 & 15.43 & ab & 0.67619 & 2451582.7058 &  9.4 & 2 & 1 &  217 &  \nodata & 20 &   93 & -2.52 \\
161 & 173.844930 & -0.895030 & 15.40 & ab & 0.60699 & 2451255.6642 &  9.3 & 2 & 2 &   18 &  7 & 15 & -100 & -1.71 \\
\enddata
\tablenotetext{a}{The complete version of this table is in the electronic
edition of the Journal.  The printed edition contains only a sample.}
\label{tab-velocities}
\end{deluxetable}

\begin{deluxetable}{cccccc}
\tablecolumns{6}
\tablewidth{0pc}
\tablecaption{Blue Horizontal Branch from WIYN Observations}
\tablehead{
\colhead {ID} & \colhead{$\alpha$} & \colhead{$\delta$} & 
\colhead{D} & \colhead{$V_{gsr}$} & \colhead{$\sigma_{V_{gsr}}$} \\
\colhead{} & \colhead{(deg)} & \colhead{(deg)} & \colhead{(kpc)} & \colhead{(\kms)} &
\colhead{(\kms)} \\
}
\startdata
Q245648 &  190.865875 & -0.926903 &   9.8  &    188  &   8 \\
Q211556 &  180.395035 & -1.820980 &   9.3  &     -6  &  13 \\
Q295219  & 188.952927 & -2.765785 &  12.0  &     76  &   8 \\ 
\enddata
\label{tab-bhb}
\end{deluxetable}

\end{document}